\documentclass[
reprint,
superscriptaddress,
amsmath,amssymb,
aps,
prd,
floatfix,
nobibnotes,
nofootinbib,
]{revtex4-2}

\usepackage{graphicx}
\usepackage{dcolumn}
\usepackage{bm}
\usepackage[english]{babel}
\usepackage[utf8]{inputenc}
\usepackage[T1]{fontenc}
\usepackage{microtype}
\usepackage[table,dvipsnames]{xcolor}
\definecolor{myblue}{RGB}{62,73,173}
\definecolor{myred}{RGB}{216, 28, 56} 
\usepackage[
colorlinks=true,
breaklinks=true,
hyperfootnotes=true,
allcolors=myblue
]{hyperref}
\usepackage[nameinlink]{cleveref}

\Crefname{section}{Sec.}{Sec.}
\Crefname{figure}{Fig.}{Figs.}
\Crefname{table}{Tab.}{Tabs.}

\usepackage{lineno}
\usepackage{physics}
\usepackage[separate-uncertainty=true]{siunitx}

\AtBeginDocument{\RenewCommandCopy\qty\SI}

\usepackage{booktabs}
\usepackage{subfigure}
\usepackage{multirow}
\usepackage{threeparttable}
\usepackage{makecell}
\usepackage{orcidlink}

\usepackage{acro}
\acsetup{
barriers/use = true,
barriers/reset = true,
}
\DeclareAcroEnding{possessive}{'s}{'s}
\NewAcroCommand\acg{m}{\acropossessive\UseAcroTemplate{first}{#1}}
\NewAcroCommand\acsg{m}{\acropossessive\UseAcroTemplate{short}{#1}}
\NewAcroCommand\aclg{m}{\acropossessive\UseAcroTemplate{long}{#1}}
\NewAcroCommand\acfg{m}{%
    \acrofull
    \acropossessive
    \UseAcroTemplate{first}{#1}%
}
\NewAcroCommand\iacsg{m}{%
    \acroindefinite
    \acropossessive
    \UseAcroTemplate{short}{#1}%
}

\DeclareAcronym{AUC}{
    short = AUC,
    long  = area under the curve
}
\DeclareAcronym{AI}{
    short = AI,
    long = artificial intelligence
}
\DeclareAcronym{BH}{
    short = BH,
    long  = black hole
}
\DeclareAcronym{BHB}{
    short = BHB,
    long  = black hole binary,
    long-plural-form = black hole binaries
}
\DeclareAcronym{BBH}{
    short = BBH,
    long  = binary black hole
}
\DeclareAcronym{CNN}{
    short = CNN,
    long  = convolutional neural network
}
\DeclareAcronym{CV}{
    short = CV,
    long  = computer vision
}

\DeclareAcronym{CNF}{
    short = CNF,
    long  = continuous normalizing flow
}

\DeclareAcronym{FMPE}{
    short = FMPE,
    long  = flow-matching parameter estimation
}
\DeclareAcronym{PE}{
    short = PE,
    long  = parameter estimation
}
\DeclareAcronym{NPE}{
    short = NPE,
    long  = neural parameter estimation
}

\DeclareAcronym{LSTM}{
    short = LSTM,
    long  = long short-term memory
}

\DeclareAcronym{JS}{
    short = JS,
    long  = Jensen-Shannon,
    long-plural-form = Jensen-Shannon divergences
}

\DeclareAcronym{DNN}{
    short = DNN,
    long  = deep neural network
}

\DeclareAcronym{DL}{
    short = DL,
    long  = deep learning
}
\DeclareAcronym{ESA}{
    short = ESA,
    long  = European Space Agency
}
\DeclareAcronym{EMRI}{
    short = EMRI,
    long  = extreme-mass-ratio inspiral
}
\DeclareAcronym{FPR}{
    short = FPR,
    long  = false positive rate
}
\DeclareAcronym{GB}{
    short = GB,
    long  = galactic binary,
    long-plural-form = galactic binaries
}
\DeclareAcronym{GR}{
    short = GR,
    long  = general relativity
}
\DeclareAcronym{GW}{
    short = GW,
    long  = gravitational wave
}

\DeclareAcronym{LDC}{
    short = LDC,
    long  = LISA Data Challenge
}
\DeclareAcronym{LIGO}{
    short = LIGO,
    long  = \href{http://www.ligo.caltech.edu/}{Laser Interferemeter Gravitational Wave Observatory}
}
\DeclareAcronym{LISA}{
    short = LISA,
    long  = \href{https://www.lisamission.org/}{Laser Interferometer Space Antenna}
}

\DeclareAcronym{MBH}{
    short = MBH,
    long  = massive black hole
}
\DeclareAcronym{MBHB}{
    short = MBHB,
    long  = massive black hole binary,
    long-plural-form = massive black hole binaries
}

\DeclareAcronym{MCMC}{
    short = MCMC,
    long  = Markov-chain Monte Carlo
}
\DeclareAcronym{MLDC}{
    short = MLDC,
    long  = \href{http://astrogravs.nasa.gov/docs/mldc/}{Mock LISA Data Challenge}
}
\DeclareAcronym{MLP}{
    short = MLP,
    long  = multi-layer perceptron
}
\DeclareAcronym{NK}{
    short = NK,
    long  = numerical kludge
}

\DeclareAcronym{NLP}{
    short = NLP,
    long  = natural language processing
}
\DeclareAcronym{OMS}{
    short = OMS,
    long  = optical metrology system
}
\DeclareAcronym{PSD}{
    short = PSD,
    long  = power spectral density
}
\DeclareAcronym{PTA}{
    short = PTA,
    long  = pulsar timing array
}
\DeclareAcronym{EPTA}{
    short = EPTA,
    long  = European pulsar timing array
}
\DeclareAcronym{PPTA}{
    short = PPTA,
    long  = Parkes pulsar timing array
}
\DeclareAcronym{CPTA}{
    short = CPTA,
    long  = Chinese pulsar timing array
}

\DeclareAcronym{NANOGrav}{
    short = NANOGrav,
    long  = North American Nanohertz Observatory for Gravitational Waves
}

\DeclareAcronym{NG}{
    short = NG,
    long  = NANOGrav
}

\DeclareAcronym{ROC}{
    short = ROC,
    long  = receiver operating characteristic
}
\DeclareAcronym{SGWB}{
    short = SGWB,
    long  = stochastic gravitational wave background
}
\DeclareAcronym{SMBH}{
    short = SMBH,
    long  = super-massive black hole
}
\DeclareAcronym{SNR}{
    short = SNR,
    long  = signal-to-noise ratio
}

\DeclareAcronym{TDI}{
    short = TDI,
    long  = time delay interferometry
}
\DeclareAcronym{TPR}{
    short = TPR,
    long  = true positive rate
}

\DeclareAcronym{VGB}{
    short = VGB,
    long  = verification galactic binary,
    long-plural-form = verification galactic binaries
}

\DeclareAcronym{CURN}{
    short = CURN,
    long  = common uncorrelated red noise,
    long-plural-form = common uncorrelated red noises
}

\DeclareAcronym{HD}{
    short = HD,
    long  = Hellings-Downs,
    long-plural-form = Hellings-Downs correlations
}

\DeclareAcronym{ODE}{
    short = ODE,
    long  = ordinary differential equation,
}

\DeclareAcronym{CDF}{
    short = CDF,
    long  = cumulative distribution function,
}

\graphicspath{{fig/}}

\renewcommand{\sec}[1]{\textbf{\emph{#1.---}}}
\newcommand{\btheta}{\boldsymbol{\theta}}

\begin{document}

\title{Accelerating Stochastic Gravitational Wave Backgrounds \\ Parameter Estimation in Pulsar Timing Arrays with Flow Matching}

\author{Bo Liang}
 \altaffiliation[]{These authors contributed equally to the work.}
 \affiliation{Center for Gravitational Wave Experiment, National Microgravity Laboratory, Institute of Mechanics, Chinese Academy of Sciences, Beijing 100190, China}
 \affiliation{Shanghai Institute of Optics and Fine Mechanics, Chinese Academy of Sciences, Shanghai 201800, China}
 \affiliation{Taiji Laboratory for Gravitational Wave Universe (Beijing/Hangzhou), University of Chinese Academy of Sciences (UCAS), Beijing 100049, China}

\author{Chang Liu}
 \altaffiliation[]{These authors contributed equally to the work.}
 \affiliation{Center for Gravitational Wave Experiment, National Microgravity Laboratory, Institute of Mechanics, Chinese Academy of Sciences, Beijing 100190, China}
 \affiliation{National Space Science Center, Chinese Academy of Sciences, Beijing 100190, China}

\author{Tianyu Zhao}
 \email{zhaotianyu@imech.ac.cn}
 \affiliation{Center for Gravitational Wave Experiment, National Microgravity Laboratory, Institute of Mechanics, Chinese Academy of Sciences, Beijing 100190, China}

\author{Minghui Du}
 \affiliation{Center for Gravitational Wave Experiment, National Microgravity Laboratory, Institute of Mechanics, Chinese Academy of Sciences, Beijing 100190, China}

\author{Manjia Liang}
 \affiliation{Center for Gravitational Wave Experiment, National Microgravity Laboratory, Institute of Mechanics, Chinese Academy of Sciences, Beijing 100190, China}

\author{Ruijun Shi}
 \affiliation{School of Physics and Astronomy, Beijing Normal University, Beijing, 100875, China}
 \affiliation{Institute for Frontiers in Astronomy and Astrophysics, Beijing Normal University, Beijing, 102206, China}

\author{Hong Guo}
 \affiliation{Escola de Engenharia de Lorena, Universidade de São Paulo, 12602-810, Lorena, SP, Brazil}

\author{Yuxiang Xu}
 \affiliation{Center for Gravitational Wave Experiment, National Microgravity Laboratory, Institute of Mechanics, Chinese Academy of Sciences, Beijing 100190, China}
 \affiliation{Shanghai Institute of Optics and Fine Mechanics, Chinese Academy of Sciences, Shanghai 201800, China}
 \affiliation{Taiji Laboratory for Gravitational Wave Universe (Beijing/Hangzhou), University of Chinese Academy of Sciences (UCAS), Beijing 100049, China}

\author{Li-e Qiang}
 \affiliation{Center for Gravitational Wave Experiment, National Microgravity Laboratory, Institute of Mechanics, Chinese Academy of Sciences, Beijing 100190, China}
 \affiliation{National Space Science Center, Chinese Academy of Sciences, Beijing 100190, China}

\author{Peng Xu}
 \email{xupeng@imech.ac.cn}
 \affiliation{Center for Gravitational Wave Experiment, National Microgravity Laboratory, Institute of Mechanics, Chinese Academy of Sciences, Beijing 100190, China}
 \affiliation{Key Laboratory of Gravitational Wave Precision Measurement of Zhejiang Province, Hangzhou Institute for Advanced Study, UCAS, Hangzhou 310024, China}
 \affiliation{Taiji Laboratory for Gravitational Wave Universe (Beijing/Hangzhou), University of Chinese Academy of Sciences (UCAS), Beijing 100049, China}
 \affiliation{Lanzhou Center of Theoretical Physics, Lanzhou University, Lanzhou 730000, China}

\author{Wei-Liang Qian}
 \affiliation{Escola de Engenharia de Lorena, Universidade de São Paulo, 12602-810, Lorena, SP, Brazil}
 
\author{Ziren Luo}
 \affiliation{Center for Gravitational Wave Experiment, National Microgravity Laboratory, Institute of Mechanics, Chinese Academy of Sciences, Beijing 100190, China}
 \affiliation{Key Laboratory of Gravitational Wave Precision Measurement of Zhejiang Province, Hangzhou Institute for Advanced Study, UCAS, Hangzhou 310024, China}
 \affiliation{Taiji Laboratory for Gravitational Wave Universe (Beijing/Hangzhou), University of Chinese Academy of Sciences (UCAS), Beijing 100049, China}

\begin{abstract}
\Acp{PTA} are essential tools for detecting the \ac{SGWB}, but their analysis faces significant computational challenges. Traditional methods like \ac{MCMC} struggle with high-dimensional parameter spaces where noise parameters often dominate, 
while existing deep learning approaches have so far been validated on synthetic datasets or require training on the full pulsar set, incurring substantial computational and memory costs.
We propose a flow-matching-based \ac{CNF} for efficient \ac{PTA} parameter estimation. Using ten pulsars selected according to published NANOGrav 12.5-year dropout factors and applied to the NANOGrav 15-year residuals, our method produces \ac{SGWB} posteriors consistent with a reference \ac{MCMC} analysis, with Jensen-Shannon divergences below \(10^{-2}\) nat. After amortized training, posterior generation is reduced from approximately 50 hours for the reference \ac{MCMC} pipeline to approximately 4 minutes for \ac{CNF} sampling. The present study demonstrates that flow-matching-based \acp{CNF} can serve as an efficient posterior-sampling accelerator for reduced \ac{PTA} datasets. Rather than providing new astrophysical constraints, the method is intended to complement conventional \ac{MCMC} analyses and to provide a scalable route toward faster inference in future \ac{PTA} applications.
\end{abstract}

\maketitle

\acbarrier
\begin{figure*}[ht!]
  \centering
  \includegraphics[width=1.0\textwidth]{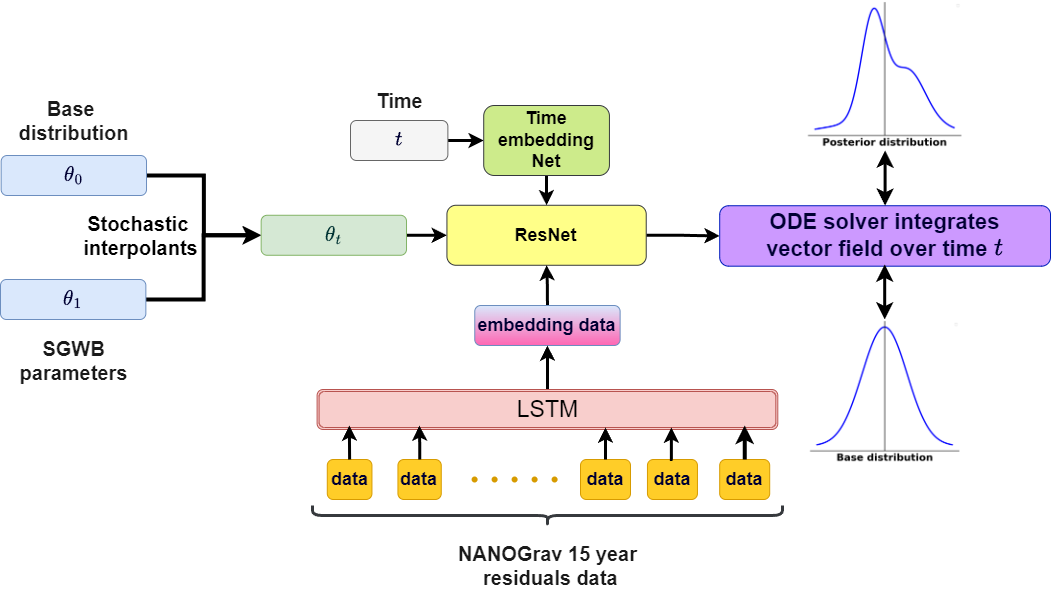}
  \caption{
    The figure illustrates the framework for training using CNFs.
\( t \) represents the sampling time, with a range from 0 to 1.
In describing the transition from the initial Gaussian distribution \( \theta_0 \) to the SGWB posterior distribution \( \theta_1 \), we use linear interpolation to construct the distribution \( \theta_t \) at any given time \( t \).
Additionally, we utilize LSTM to construct an embedding network, aimed at processing NG15 real PTA residual data related to the power-law spectrum of HD.
Once CNFs are trained, they can use an ODE solver to transform the initial base distribution into posterior distribution. This process involves utilizing the invertible property of CNFs to solve a mapping obtained from a neural ordinary differential equation (ODE), thereby achieving a transition from a simple distribution to a complex one.
  }
  \label{fig:model}
\end{figure*}


\sec{Introduction} \Acp{PTA} have emerged as a crucial tool in astrophysics, offering unique insights into the cosmos. Recent observations from various collaborations, including the \ac{NANOGrav}~\cite{McLaughlin_2013,brazier2019nanogravprogramgravitationalwaves}, \ac{EPTA}~\cite{Ferdman2010TheEP}, \ac{PPTA}~\cite{Manchester2006ThePP}, and \ac{CPTA}~\cite{Xu2023SearchingFT}, have confirmed the existence of \ac{SGWB}~\cite{Agazie2023TheN1, Antoniadis2023TheSD}. This groundbreaking achievement highlights the potential of \acp{PTA} to probe the fabric of spacetime, enriching our understanding of the evolutions of supermassive black holes and other cosmological phenomena~\cite{addazi_have_2024}. Such capabilities provide a promising avenue for unraveling mysteries that lie beyond the reach of traditional observational methods \cite{taylor_nanohertz_2021}.

Despite these advancements, traditional data analysis methods—particularly \ac{MCMC} techniques—face significant challenges in parameter estimation for \acp{PTA}~\cite{babak_forecasting_2024,van_haasteren_measuring_2009,enterprise,justin_ellis_2017_1037579}. The high dimensionality of the parameter space and the extended data length increase the challenges of both time-domain and frequency-domain analyses \cite{taylor_searching_2013}. \ac{MCMC} methods are often hindered by slow convergence and require extensive computational resources, rendering them inefficient when confronted with the increasingly complex datasets produced by modern \acp{PTA} \cite{falxa_modeling_2024,freedman_efficient_2023}. Moreover, intricate interdependencies among parameters can lead to degeneracy \cite{van_haasteren_new_2014}, resulting in biased estimates and significantly increased computational costs. As \acp{PTA} datasets continue to grow in size and complexity, these limitations underscore the urgent need for more efficient and robust estimation techniques  \cite{hobbs_international_2010}.

Artificial-intelligence-based posterior estimators are becoming increasingly useful for scientific data analysis, particularly in the context of \ac{PE} for \ac{PTA} studies and \ac{GW} detection, as seen with \ac{LIGO} and \ac{LISA} \cite{Dax2021RealtimeGS, Dax2022NeuralIS, Liang_2024}. Deep learning has shown promise in enhancing data analysis capabilities; however, existing approaches, such as \ac{NPE} \cite{papamakarios2018fastepsilonfreeinferencesimulation, lueckmann2017flexiblestatisticalinferencemechanistic, greenberg2019automaticposteriortransformationlikelihoodfree}, often fall short due to the still slower processing times and potential inaccuracies. These limitations become particularly evident when applied to real observational scenarios in \ac{PTA} data analysis and \ac{SGWB} detection, where the intricate nature of signals demands more sophisticated techniques~\cite{Dax2023FlowMF}. 
Previous studies have explored \ac{NPE}-based and related approaches for \ac{PTA} data and \ac{SGWB} detection \cite{Shih2023FastPI,vallisneri_rapid_2024}. \citet{Shih2023FastPI} demonstrated rapid \ac{NPE} inference using 12.5 years of synthetic data. \citet{vallisneri_rapid_2024} first present a variational-Bayes framework and trained model on all pulsars in \acs{NG}15 dataset under two cases: modeling only \ac{CURN} and jointly modeling red noise with \ac{HD}–correlated \ac{SGWB} parameters.

We introduce a flow-matching-based \ac{CNF} method for parameter estimation in \ac{PTA} data~\cite{chen2019neuralordinarydifferentialequations,lipman2023flowmatchinggenerativemodeling,Dax2023FlowMF,Liang_2024}. Utilizing recent advancements in flow-matching techniques~\cite{Dax2023FlowMF, Liang_2024, Gebhard2024FlowMF}, our approach is trained on a reduced set of 10 pulsars out of 68, selected according to published dropout factors from the \ac{NG}12.5 analysis~\cite{arzoumanian_nanograv_2020}. Applying our method to the \ac{NG}15 dataset~\cite{Agazie2023TheN1}, we perform posterior estimation on real data characterized by the \ac{HD}-correlated power-law spectrum of the \ac{SGWB}~\cite{Hellings:1983fr}. Our results demonstrate that the obtained \ac{SGWB} posterior distributions are consistent with those from a reference \ac{MCMC} analysis~\cite{enterprise, justin_ellis_2017_1037579, robert2020markovchainmontecarlo}, while significantly reducing the posterior-generation time after amortized training. We believe this advancement can provide a useful complement to conventional \ac{PTA} inference pipelines and may facilitate faster exploratory and repeated posterior analyses for future \ac{PTA} datasets.

\sec{Dataset Generation} In constructing our simulated pulsar timing dataset, our goal is to accurately model the key sources of signals and noise relevant to our parameter estimation method. Based on the 15 years of \ac{NG} data~\cite{Agazie2023TheN1}, we generated 1.5 million pulsar timing residual time series. We focused on ten pulsars selected according to their dropout factors for common uncorrelated red noise in the \ac{NG}12.5 dataset, as detailed in \Cref{table:10pulsar}. We acknowledge that this selection---while well motivated by \ac{NG}12.5 analysis---has not been re-evaluated on the \ac{NG}15 data, and to the best of our knowledge, a corresponding dropout study for \ac{NG}15 is not yet publicly available. We therefore emphasize that the ten-pulsar subset is used as a reduced-data demonstration of the proposed inference framework. The present analysis should not be interpreted as a replacement for a full-array \ac{NG}15 production analysis. Instead, it provides a controlled setting in which the \ac{FMPE} posterior can be directly compared with a conventional \ac{MCMC} reference posterior for the same reduced dataset.

\begin{table}[ht!]
    \centering
    \footnotesize
    \caption{The ten pulsars providing the strongest evidence for \acs{SGWB} and their high dropout factors~\cite{arzoumanian_nanograv_2020}}
    \label{table:10pulsar}
    \begin{tabular*}{\columnwidth}{@{\extracolsep{\fill}} l c @{\hspace{0.2em}}  @{\hspace{0.2em}} l c @{}}
        \toprule
        \hline
        \textbf{Pulsar} & \textbf{Dropout Factor} & \textbf{Pulsar} & \textbf{Dropout Factor} \\
        \midrule
        J1909$-$3744 & 17.6 & J0030$+$0451 & 2.4 \\
        J2317$+$1439 & 14.5 & J1910$+$1256 & 2.4 \\
        J2043$+$1711 & 6.0 & J1744$-$1134 & 2.5 \\
        J1600$-$3053 & 5.3 & J1944$+$0907 & 3.3 \\
        J1918$-$0642 & 3.4 & J0613$-$0200 & 3.4 \\
        \hline
        \bottomrule
    \end{tabular*}
\end{table}

\begin{figure}[ht!]
  \centering
  \includegraphics[width=0.47\textwidth]{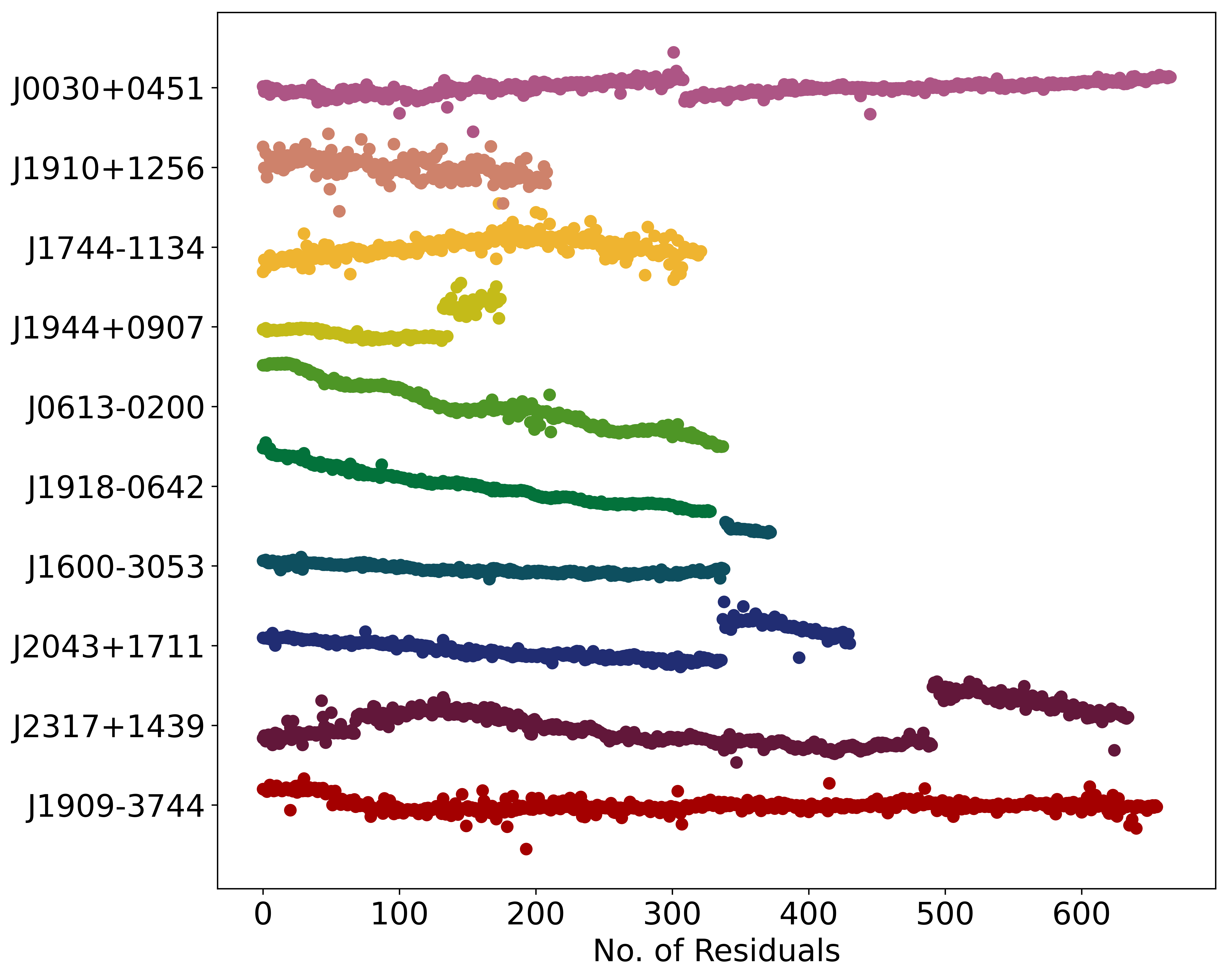}
  \caption{ 
  We selected real residual data from ten pulsars in \ac{NG}15 for analysis. These pulsars provided strong evidence for a SGWB in \ac{NG}12.5 and define the reduced dataset used in this methodological demonstration.
  }
  \label{fig:resdual}
\end{figure}

The \ac{SGWB} signal is modeled as a common process affecting all pulsars, characterized by an amplitude \( A_{\mathrm{GW}} \) and a spectral index \( \gamma_{\mathrm{GW}} \). The \ac{PSD} of the \ac{SGWB}-induced timing residuals follows a power-law spectrum given by
\begin{equation}
    P_{\mathrm{SGWB}}(f) = \frac{A_{\mathrm{GW}}^2}{12\pi^2} \left( \frac{f}{f_{\mathrm{ref}}} \right)^{-\gamma_{\mathrm{GW}}}f^{-3}_{\mathrm{ref}},
\end{equation}
where \( f \) is the frequency 
and \( f_{\mathrm{ref}} = 1\, \mathrm{yr}^{-1} \)
is the reference frequency. To capture the spatial correlations between pulsars due to the \ac{SGWB}, we incorporated the \ac{HD}-correlation matrix \( \chi_{IJ} \), defined by
\begin{equation}
\begin{split}
    \chi_{IJ} = \frac{3}{2} \Bigg[ \left( \frac{1 - \cos \theta_{IJ}}{2} \right) \ln  & \left( \frac{1 - \cos \theta_{IJ}}{2} \right) \\
    & - \frac{1}{6} \left(1 - \cos \theta_{IJ}\right) + \frac{1}{3} \Bigg],
\end{split}
\end{equation}
where \( \theta_{IJ} \) is the angular separation between pulsars \( I \) and \( J \). We simulated \( \chi_{IJ} \) by calculating the angular separations between the ten selected pulsars and evaluating the HD function for each pair. Each pulsar also has its own intrinsic red noise, with amplitude \( A_r^{(i)} \) and spectral index \( \gamma_r^{(i)} \), following a similar power-law spectrum:
\begin{equation}
    P_{\mathrm{red}}^{(i)}(f) = \frac{A_r^{(i)\,2}}{12\pi^2} \left( \frac{f}{f_{\mathrm{ref}}} \right)^{-\gamma_r^{(i)}}f^{-3}_{\mathrm{ref}}.
\end{equation}
Specifically, for each pulsar, the red noise amplitude \(\log_{10} A_r^{(i)}\) was drawn uniformly from the range \([-19, -13]\), and the spectral index \(\gamma_r^{(i)}\) was sampled uniformly from \([1, 7]\).

In order to better approximate the white-noise level in the training simulations, we processed the white-noise level $\sigma$ for each pulsar based on its time-of-arrival uncertainty array \( \mathrm{ToA\_Err}_i \). Specifically, we set \( \sigma = \sigma_i \), where the effective white-noise level for each selected pulsar is given by \( \sigma_i = \mathrm{mean}(\mathrm{ToA\_Err}_i) \) after processing the \ac{NG}15 data with the \texttt{ENTERPRISE} software. The average level of these values is roughly between 100 and 500 ns. This treatment is a simplified approximation of the full \ac{PTA} white-noise model. Standard \ac{PTA} analyses can include backend-dependent EFAC/EQUAD/ECORR-like terms, and incorporating these additional white-noise parameters into the simulator would be necessary for a production-level full-array analysis. In the present work, this approximation is used to construct a controlled simulation set for training the neural posterior estimator, while the reference posterior is still obtained from a conventional \ac{MCMC} pipeline for the same reduced dataset.

Before training, we pre-processed the simulated residuals to optimize the performance of our machine learning model. This involved rescaling the residuals to ensure that they were within a suitable range for neural network training. Specifically, we multiply every entry \(r\) by \(10^7\) and then clip the values to lie within \([-1000, +1000]\). 

The ten pulsars contribute a total of 4,944 projected residuals, providing a rich dataset for analysis. We partitioned the 1.5 million generated time series into two sets: 98\% were used for training the \acp{CNF}, and the remaining 2\% were reserved for model validation. This dataset enabled us to effectively train and validate our model.

\begin{figure}[ht!]
  \centering
  \includegraphics[width=0.47\textwidth]{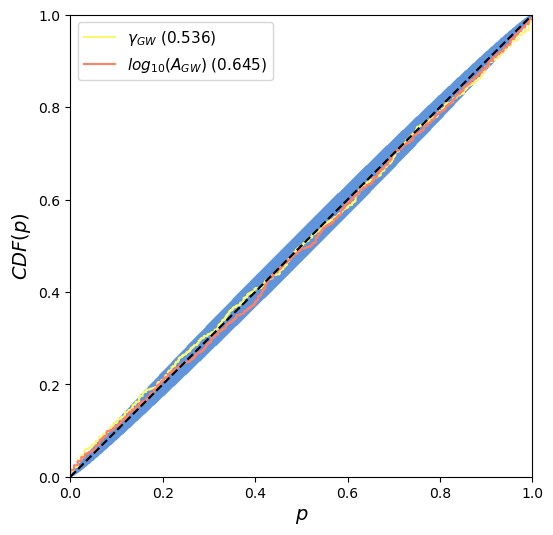}
  \caption{
   In the P-P plot of 800 injections, the p-values for each parameter are indicated in the legend, and the average p-value across all parameters is calculated to be 0.591.
  }
  \label{fig:pp}
\end{figure}

\begin{figure}[ht!]
  \centering
  \includegraphics[width=0.47\textwidth]{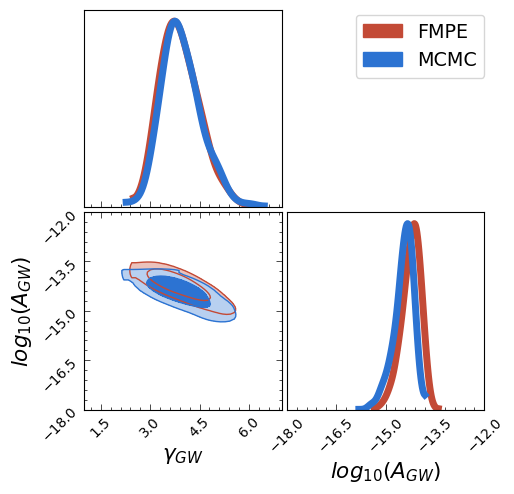}
  \caption{
  Comparison between the \ac{FMPE} posterior and the reference \ac{MCMC} posterior for the \ac{HD}-correlated power-law \ac{SGWB} parameters in the reduced ten-pulsar dataset. The two methods give consistent marginal distributions and correlations for \(\gamma_{\mathrm{GW}}\) and \(\log_{10} A_{\mathrm{GW}}\).
  }
  \label{fig:corner}
\end{figure}

\sec{Model Architecture} 
In this Letter, we employ \ac{CNF} to fit the posterior distribution of the \ac{NG}15 dataset~\cite{Agazie2023TheN1}, modeling it with a power-law spectrum and incorporating \ac{HD} correlations. To optimize the \acp{CNF}, we utilize flow matching~\cite{lipman2023flowmatchinggenerativemodeling, Liang_2024, Dax2023FlowMF}, a novel technique in artificial intelligence that accelerates the training of \acp{CNF}. This method, known as \ac{FMPE}, enhances efficiency and precision in modeling complex distributions.

A \ac{CNF} defines a transformation of a probability density through a continuous-time dynamical system. Given a base distribution \( p_0(\boldsymbol{\theta}_0) \) and a target distribution \( p_1(\boldsymbol{\theta}_1) \), the transformation is defined by the \ac{ODE}~\cite{chen2019neuralordinarydifferentialequations}:
\begin{equation}
    \frac{d}{dt} \psi_{t,x}(\btheta) = v_{t,x}(\psi_{t,x}(\btheta)).
\end{equation}
Where, \( \psi_{t,x}(\btheta) \) describes the transition from the base distribution \( p_0(\boldsymbol{\theta}_0) \) to the target distribution  \( p_1(\boldsymbol{\theta}_1) \).

To train the CNF, we employ the flow-matching loss function \cite{tong2024improvinggeneralizingflowbasedgenerative, lipman2023flowmatchinggenerativemodeling}, which minimizes the difference between the vector field \( v_{t,x}(\psi_{t,x}(\btheta)) \) and the optimal vector field \( u_t(\btheta|\btheta_1) \) that transports \( p_0 \) to \( p_1 \). 
For this work, we define the Gaussian probability paths \( p_{t}(\btheta|\btheta_{1}) \) as:
\begin{equation}
    p_t(\theta|\theta_1) = \mathcal{N}\left(\theta_1t,(1-t\right)^2 \cdot I_d) ,
\end{equation}
The probability path under consideration is capable of transforming a d-dimensional standard normal distribution at $t = 0$ into a Gaussian distribution with a centre at $\theta_{1}$ at $t = 1$.
Thus the optimal vector field \(u_t(\btheta|\btheta_1)\) is defined as \cite{liu2022flowstraightfastlearning, Albergo2022BuildingNF}:
\begin{equation}
u_t(\btheta|\btheta_1) =\btheta_1 - \btheta_{0}. 
\end{equation}
When averaging over many $\theta_{1} \sim P(\theta)$, regressing onto these target vector fields $ u_{t}(\theta|\theta_{1})$ results in a vector field $v_{t,x}(\theta)$ that transforms the base into the correct target distribution. Thus, the loss function we need to minimize is:
\begin{equation}
\mathcal{L_{\mathrm{FM}}} = \mathbb{E}_{t \sim p(t)}  
\mathbb{E}_{ \theta_1 \sim p(\theta)}
\mathbb{E}_{x \sim p(x|\theta_1)}
\mathbb{E}_{ \theta_t \sim p_t(\theta_t|\theta_1)}
\parallel r \parallel^2, \ \ 
\end{equation}
\begin{equation}
r = v_{t,x}(\theta_t) - u_t(\theta_t|\theta_1).
\end{equation}

In \ac{PTA} research, a key challenge lies in effectively addressing the high-dimensional noise parameters, which can span tens of dimensions, compared to the two-dimensional SGWB parameters associated with the \ac{HD} correlation. The dominance of noise parameter dimensions often limits the focus on \ac{SGWB} signals, leading to high computational costs and convergence difficulties in \ac{MCMC} analysis. 

To tackle these challenges, we introduce a signal-aware loss reweighting strategy with a tunable hyperparameter, defined as:  
\begin{equation}
\mathcal{L} = \lambda \cdot \mathcal{L}^{\mathrm{SGWB}}_{\mathrm{FM}} + (1-\lambda) \cdot \mathcal{L}^{\mathrm{Noise}}_{\mathrm{FM}}, 
\end{equation} 
\begin{equation}
 \mathcal{L}^{\mathrm{SGWB}}_{\mathrm{FM}} = \ \parallel v_{t,x}(\theta_t) - u_t(\theta_t|\theta_1) \parallel^2, \theta_{1} \sim p(\theta)_{\mathrm{SGWB}}, 
\end{equation}
\begin{equation}
\mathcal{L}^{\mathrm{Noise}}_{\mathrm{FM}}  = \ \parallel v_{t,x}(\theta_t) - u_t(\theta_t|\theta_1) \parallel^2, \theta_{1} \sim p(\theta)_{Noise}. 
\end{equation}
where $p(\theta)_{\mathrm{SGWB}}$ represents the true vector field  sampled from the two SGWB parameters and $p(\theta)_{Noise} $ represents the true vector field, sampled from the 20 red noise. 
This approach enables flexible reweighting of the model’s attention, allowing it to prioritize the fitting of the \ac{SGWB}. By balancing the contributions of \ac{SGWB} and noise parameters, our method improves both the efficiency and accuracy of parameter estimation.

Our model architecture comprises two main components, as illustrated in \Cref{fig:model}. The \textbf{first component} is an embedding network designed to process the 4,944 residuals from the ten pulsars and perform feature extraction. Since our flow network ultimately fits a two-dimensional vector field, the embedding network is essential for reducing the high-dimensional input data to a manageable size. We employ a \ac{LSTM} network as the embedding network~\cite{1997Long}, which compresses the features of each pulsar into 100 dimensions. These features are further refined and compressed into 50 dimensions using a residual network~\cite{he2015deepresiduallearningimage}. The resulting compressed features, denoted as \( \mathbf{c} \in \mathbb{R}^{50} \), serve as conditional inputs to the flow network.

The \textbf{second component} is the flow network itself, consisting of 56 residual blocks with gradually decreasing sizes. The vector field \( \mathbf{v}_{t,x}(\boldsymbol{\theta}_t) \) is parameterized by the neural network \( f_{\boldsymbol{\theta}} \) and conditioned on the embedded features \( \mathbf{c} \) and the time parameter \( t \). Specifically, the flow network takes as input the triplet \( (\boldsymbol{\theta}_t, t, \mathbf{c}) \) and outputs the vector field:
\begin{equation}
\mathbf{v}_{t,x}(\boldsymbol{\theta}_t) = f_{\boldsymbol{\theta}}(\boldsymbol{\theta}_t, t, \mathbf{c}).
\end{equation}

To implement our normalizing flow and embedding networks, we chose the  \texttt{PyTorch}~\cite{paszke2019pytorchimperativestylehighperformance} framework combined with the \texttt{numpy}~\cite{harris2020array} library, as well as the \texttt{zuko}~\cite{rozet2022zuko} and \texttt{lampe}~\cite{rozet2021lampe} software packages.
The model was trained for 100 epochs utilizing a batch size of 512, employing the Adam optimizer~\cite{kingma2017adammethodstochasticoptimization} for efficient convergence.
The initial learning rate was set to 0.0001 and was gradually reduced to zero using a cosine annealing schedule. Training was conducted on an NVIDIA RTX 4090 GPU and took approximately 20 hours. Remarkably, sampling from the trained \ac{CNF} to produce posterior distributions involves solving the ODE, which takes approximately 240 seconds for 8,192 samples using a Runge-Kutta solver~\cite{kidger2021hey, torchdiffeq}. In contrast, generating an equivalent number of posterior draws using the traditional \ac{MCMC} pipeline takes approximately 50 hours on an Apple M3 Max chip\footnote{Our ``traditional'' MCMC baseline follows the standard NG15 data analysis pipeline implemented in an x86\_64 Python 3.9 environment (run under Rosetta 2 on the M3 Max). We use \texttt{Tempo2/libstempo}~\cite{hobbs2009tempo2,vallis2025libstempo} for timing‐model interfacing and \texttt{ENTERPRISE} for likelihoods, with white noise modeled via EFAC per backend and intrinsic red noise and SGWB both represented as Fourier‐basis Gaussian processes with power‐law spectra (30 Fourier basis components for red noise, 14 for the SGWB, using the Hellings–Downs overlap reduction function). Sampling is performed with \texttt{PTMCMCSampler}~\cite{justin_ellis_2017_1037579} (5 million iterations, yielding  $\sim 5\times 10^5$ effective samples) using global and pairwise jump groups. On a single Apple M3 Max, this configuration requires $\sim$50 hours to reach convergence.}. The quoted \ac{MCMC} runtime refers to the conventional reference pipeline used to obtain the posterior for the reduced ten-pulsar dataset, including the \ac{SGWB} and intrinsic red-noise parameters. By contrast, the \ac{CNF} sampling time refers to posterior generation after amortized training. Therefore, the comparison should be understood as the cost of producing posterior samples once the neural posterior estimator has been trained, rather than as an identical end-to-end replacement of the full \ac{MCMC} workflow.


\begin{table}[ht!]
    \centering
    \footnotesize
    \caption{JS divergence of various estimates relative to the reference MCMC posterior.  (JSD values in $\times10^{-2}$\,nat)}
    \label{table:js_compare}
    \begin{tabular*}{0.95\columnwidth}{@{\extracolsep{\fill}} l cccc @{}}
        \toprule
         \hline
        \multirow{2}{*}{} & \multicolumn{3}{c}{\textbf{FMPE}} & \multirow{2}{*}{\centering \textbf{Independent MCMC}} \\
        \cmidrule(lr){2-4}
        & $\lambda=0.5$ & $\lambda=0.7$ & $\lambda=0.9$ &  \\
        \midrule
        $\gamma_{\mathrm{GW}}$ & 0.5353 & 0.5271 & 0.6730 & 0.4964 \\
        $\log_{10} A_{\mathrm{GW}}$ & 0.0103 & 0.0103 & 0.0126 & 0.0088 \\
        \hline
        \bottomrule
    \end{tabular*}
    \begin{tablenotes}
    \footnotesize
        \item [a] ``Reference MCMC'' denotes our baseline posterior. 
        \item [b] ``Independent MCMC'' is a fresh independent MCMC run on the same data, evaluated against that reference.  
  \end{tablenotes}
\end{table}

\sec{Results}
We apply our method to the \ac{NG}15 dataset (\Cref{fig:resdual}) to estimate the parameters of a \ac{SGWB} incorporating \ac{HD} correlations. \Cref{fig:corner} presents a corner plot comparing the posterior distributions obtained from our flow-matching-based \ac{CNF} model with those from the reference \ac{MCMC} analysis. The \ac{CNF}-generated posteriors closely align with the \ac{MCMC} results, capturing the correlation between the \ac{SGWB} amplitude \(\log_{10} A_{\mathrm{GW}}\) and spectral index \( \gamma_{\mathrm{GW}} \). As shown in \Cref{table:js_compare}, the \ac{JS} divergence between our method and \ac{MCMC} results is below \(10^{-2}\) nat for the \ac{SGWB} parameters. Because the neural posterior estimator is trained on simulations and applied to the real \ac{NANOGrav} residuals, simulation-to-real mismatch is an important source of possible systematic error. We therefore compare the \ac{FMPE} posterior directly with a conventional \ac{MCMC} posterior evaluated on the real data. The close overlap of the marginal contours and the \ac{JS} divergences below \(10^{-2}\) nat indicate that, for the reduced-dataset problem studied here, the simulated training distribution covers the posterior support required by the real-data inference. We note, however, that \ac{JS} divergence alone does not fully characterize posterior-tail accuracy. Therefore, we use it together with the injection-based P--P calibration test and the independent-\ac{MCMC} comparison as complementary diagnostics.

To ensure a meaningful comparison, the reference \ac{MCMC} posterior is obtained using the standard \texttt{ENTERPRISE}-based likelihood setup for the selected pulsars, with backend-dependent EFAC white-noise terms and Fourier-basis Gaussian-process descriptions of the intrinsic red noise and the \ac{HD}-correlated \ac{SGWB}. The simplified white-noise treatment used in the training simulations is therefore a modeling approximation in the amortized simulator, rather than a simplification of the \ac{MCMC} reference posterior itself. For the reduced problem considered here, posterior generation is reduced from approximately 50 hours for the reference \ac{MCMC} run to approximately 4 minutes for \ac{CNF} sampling after training.

To evaluate calibration for the target \ac{SGWB} parameters, we perform a \acl{CDF} P--P plot analysis, as depicted in \Cref{fig:pp}. The P--P plot is computed from 800 injections. The p-values are 0.536 for \(\gamma_{\mathrm{GW}}\) and 0.645 for \(\log_{10} A_{\mathrm{GW}}\), with an average p-value of 0.591. This indicates no statistically significant evidence for miscalibration of the \ac{SGWB} posterior in the tested injection set. This calibration test is focused on the \ac{SGWB} parameters, which are the primary target of the present work. The red-noise parameters are treated as nuisance parameters in this Letter, and a full posterior-predictive validation of the complete nuisance-parameter space is left for future production-level analyses.

We further checked the convergence of the reference \ac{MCMC} chains. The running means of all 22 sampled parameters, including the two \ac{SGWB} parameters, become stable after the initial transient. We also computed the split Gelman--Rubin diagnostic and found that all parameters satisfy \(\hat{R}-1<10^{-2}\), with the two \ac{SGWB} parameters also lying below this threshold. These diagnostics, shown in Appendix~\ref{app:mcmc_convergence}, support the use of the \ac{MCMC} samples as the reference posterior in the \ac{FMPE} comparison.

As \ac{PTA} datasets grow, traditional \ac{MCMC}-based techniques will face increasing computational cost. Machine-learning-based posterior estimators offer a promising complementary tool for \ac{PTA} data analysis, particularly when repeated inference over large datasets or multiple model configurations is required. The present work focuses on inference acceleration and does not claim new astrophysical constraints or new physical modeling.

\sec{Discussion}
In this study, we introduced \acp{CNF} as an efficient technique for posterior inference in a reduced \ac{PTA} data-analysis problem. The LSTM-based embedding network compresses the variable-length timing residuals into a compact representation, while the signal-aware loss reweighting prioritizes the two \ac{SGWB} parameters over the higher-dimensional nuisance red-noise sector.

The present reduced-pulsar analysis should be regarded as a proof-of-principle demonstration of accelerated inference rather than a full-array \ac{PTA} production analysis. Extending the method to a complete \ac{NG}15 analysis with an \ac{NG}15-optimized pulsar selection, or ideally with all available pulsars, is an important direction for future work. A natural extension of the present framework is to augment the simulator with a more complete \ac{PTA} noise prescription, including backend-dependent white-noise parameters and other system-noise terms. This extension would not require a conceptual change of the \ac{CNF} architecture, but would enlarge the parameter vector and the simulation space to be covered during training.

The present validation also does not remove the need for more complete simulation-to-real tests in future full-array analyses. In particular, production applications should include more detailed white-noise and system-noise prescriptions, and should verify posterior coverage under these enlarged simulation models. Although the \ac{SGWB} posterior passes the \ac{JS}-divergence, independent-\ac{MCMC}, and P--P calibration checks considered here, these diagnostics do not constitute a complete validation of all nuisance-parameter tails. Future full-array applications should include posterior-predictive checks and simulation-based calibration over the enlarged noise-parameter space.

Machine-learning-based posterior estimators offer a promising complementary tool for \ac{PTA} data analysis, particularly when repeated inference over large datasets or multiple model configurations is required. The present work focuses on inference acceleration and does not claim new astrophysical constraints or new physical modeling.

\sec{Acknowledgments}
This study is supported by the National Key Research and Development Program of China (Grant No. 2021YFC2201901, Grant No. 2021YFC2203004, Grant No. 2020YFC2200100 and Grant No. 2021YFC2201903). International Partnership Program of the Chinese Academy of Sciences, Grant No. 025GJHZ2023106GC. We also gratefully acknowledge the financial support from Brazilian agencies Funda\c{c}\~ao de Amparo \`a Pesquisa do Estado de S\~ao Paulo (FAPESP), Fundação de Amparo à Pesquisa do Estado do Rio Grande do Sul (FAPERGS), Funda\c{c}\~ao de Amparo \`a Pesquisa do Estado do Rio de Janeiro (FAPERJ), Conselho Nacional de Desenvolvimento Cient\'{\i}fico e Tecnol\'ogico (CNPq), and Coordena\c{c}\~ao de Aperfei\c{c}oamento de Pessoal de N\'ivel Superior (CAPES).

\appendix
\section{MCMC convergence diagnostics}
\label{app:mcmc_convergence}

\begin{figure}[htbp!]
  \centering
  \includegraphics[width=0.47\textwidth]{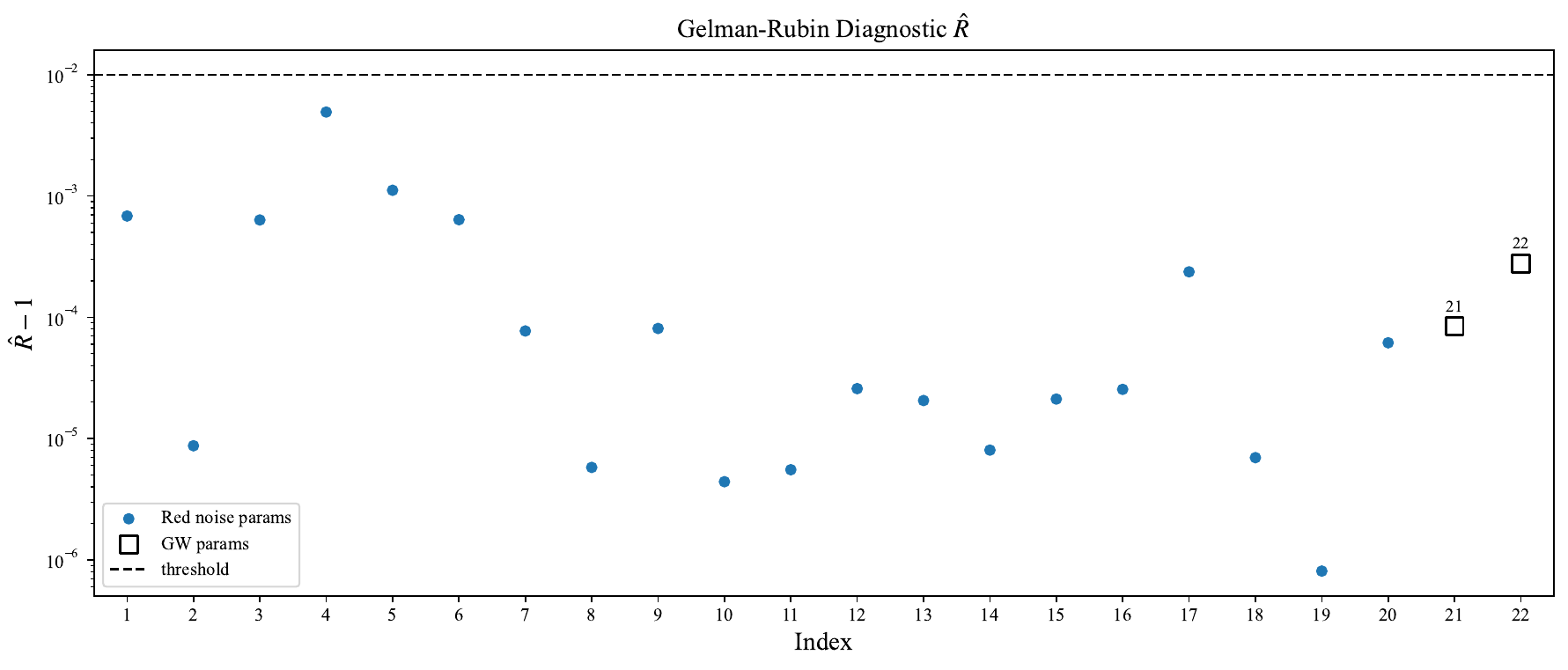}
  \caption{Split Gelman--Rubin diagnostic for the reference \ac{MCMC} chains. All red-noise and \ac{SGWB} parameters satisfy \(\hat{R}-1<10^{-2}\), indicated by the dashed threshold.}
  \label{fig:rhat}
\end{figure}

\begin{figure*}[htp!]
  \centering
  \includegraphics[width=0.95\textwidth]{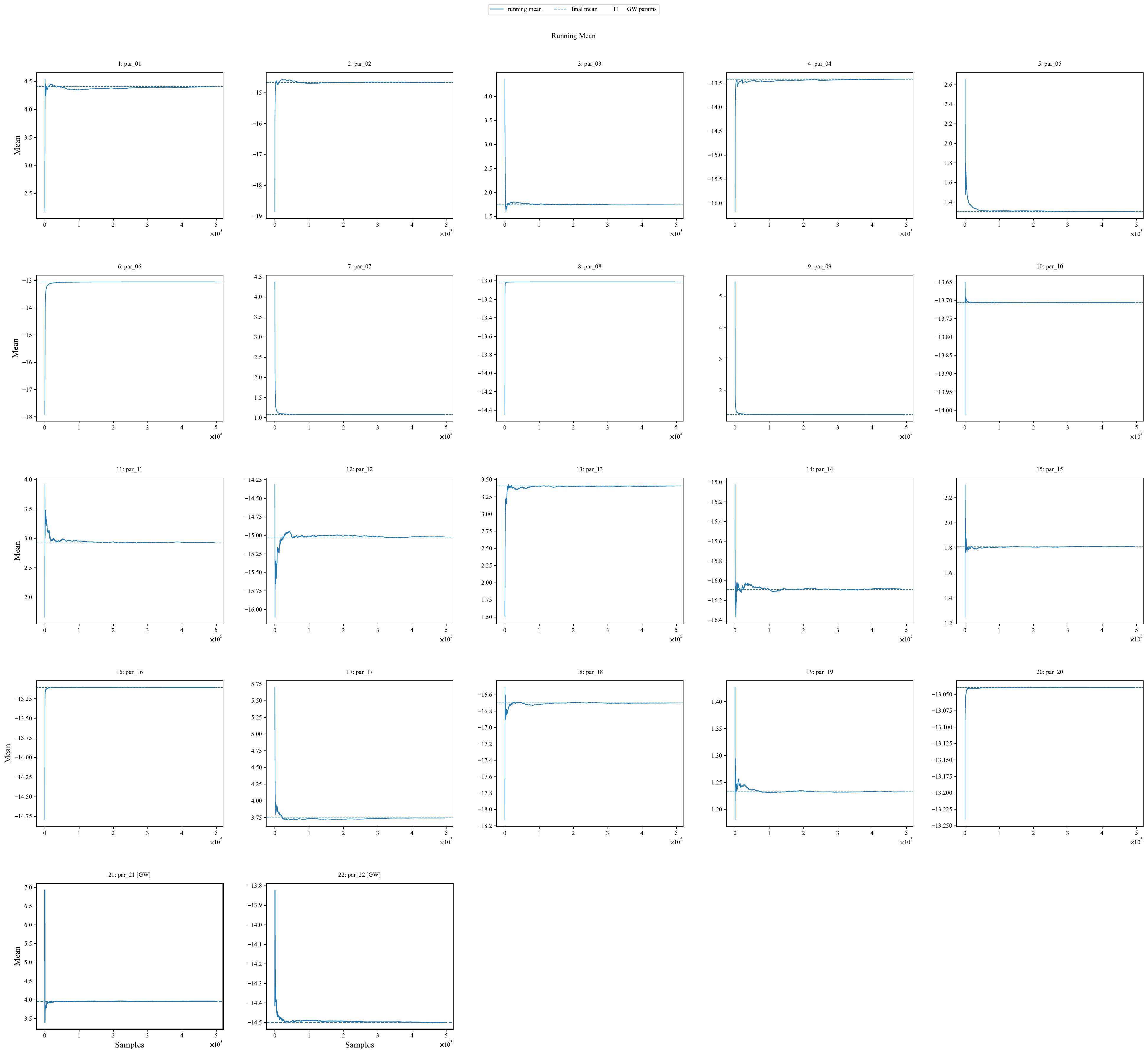}
  \caption{Running-mean diagnostic for the reference \ac{MCMC} chains. The 22 panels correspond to the red-noise parameters and the two \ac{SGWB} parameters. The running means become stable after the initial transient.}
  \label{fig:running_mean}
\end{figure*}

We include the convergence diagnostics of the reference \ac{MCMC} chains used for comparison with the \ac{FMPE} posterior. \Cref{fig:running_mean} shows the running means of the 22 sampled parameters. After the initial transient, the running means stabilize for both the red-noise parameters and the two \ac{SGWB} parameters. \Cref{fig:rhat} shows the split Gelman--Rubin diagnostic, for which all parameters satisfy \(\hat{R}-1<10^{-2}\). These diagnostics support the use of the reference \ac{MCMC} samples in the posterior comparison of the main text.

\bibliographystyle{myREVTeX4-2}
\bibliography{ref}

\end{document}